\documentclass[epj]{svjour}

\usepackage{graphicx}% Include figure files
\usepackage{dcolumn}% Align table columns on decimal point
\usepackage{bm}% bold math
\usepackage{color}
\usepackage{graphics}
\usepackage{amssymb}

\begin{document}
\title{Nonequilibrium dynamics of the three-dimensional Edwards-Anderson spin-glass 
model with Gaussian couplings: strong heterogeneities and the backbone picture}

\author{F. Rom\'a\inst{1} \and S. Bustingorry\inst{2} \and P. M. Gleiser\inst{2}}

\institute{
$^1$ Departamento de F\'{\i}sica, Universidad Nacional de 
San Luis, INFAP, CONICET, Chacabuco 917, D5700BWS San Luis, Argentina \\
$^2$ CONICET, Centro At\'omico Bariloche, R8402AGP San Carlos de Bariloche, 
R\'{\i}o Negro, Argentina}

\authorrunning{F. Rom\'a \textit{et al.}}
\titlerunning{Nonequilibrium dynamics of the three-dimensional Edwards-Anderson spin-glass model}

\date{Received: date / Revised version: \today }

\abstract{
We numerically study the three-dimensional Edwards-Anderson model with Gaussian couplings,
focusing on the heterogeneities arising in its nonequilibrium dynamics.
Results are analyzed in terms of the backbone picture, which links strong dynamical heterogeneities to spatial heterogeneities emerging from the correlation of local rigidity of the bond network.
Different two-times quantities as the flipping time distribution and the 
correlation and response functions, are evaluated over the full system and
over high- and low-rigidity regions.  
We find that the nonequilibrium dynamics of the model
is highly correlated to spatial heterogeneities.
Also, we observe a similar physical behavior to that previously found in
the Edwards-Anderson model with a bimodal (discrete) bond distribution.
Namely, the backbone behaves as the main structure that supports the spin-glass phase,
within which a sort of domain-growth process develops,
while the complement remains in a paramagnetic phase,
even below the critical temperature.
\PACS{
      {75.10.Nr}{Spin-glass and other random models}   \and
      {75.40.Gb}{Dynamic properties (dynamic susceptibility, spin waves, spin diffusion, 
                 dynamic scaling, etc.)} \and
      {75.40.Mg}{Numerical simulation studies} \and
      {75.50.Lk}{Spin glasses and other random magnets}
     } % end of PACS codes
} %end of abstract
\maketitle
%.....................................................................
\section{Introduction}

The idea that different structures with a conventional physical behavior 
are hidden inside of a disordered and frustrated material, is not new.
Historically this approach was one of the earliest considered in order to make 
a simple and intuitive picture of spin glasses \cite{Binder1986,Mydosh1993}.  
Although superparamagnetism \cite{Tholence1974,Wohlfarth1977} is a crude approximation 
of a spin-glass system, since it assumes the existence of independent 
non-interacting magnetic clusters with some kind of internal magnetic order,
for many years experimental data were interpreted within this theoretical framework.  
This was not a capricious choice: Experiments at temperatures far above 
the freezing temperature $T_f$, e. g., dc-susceptibility measurements,    
can easily be understood in these terms \cite{Morgownik1981}.    

For temperatures close or below $T_f$, the experimental evidence indicates that 
such description is still possible if a spin glass is supposedly formed by 
interacting clusters or ``building blocks'' of spins \cite{Mydosh1993}.  
Some attempts were made to explore this possibility (see Ref.~\cite{Binder1986}).  
For example, a phenomenological and intuitive picture has been proposed to explain 
several experimental results on canonical spin glasses \cite{Mydosh1993,Verbeek1980}.  
The clusters are defined as groups of spins linked by exchange bonds stronger 
than the thermal energy.  Considering RKKY interactions 
\cite{Ruderman1954,Kasuya1956,Yosida1957} between these spins, 
it is possible to delineate structures which grow as the temperature decreases.  
At $T_f$ an infinite-correlated cluster (a ``dynamic backbone'') 
emerges and percolation-like features can be identified.
  
Such simple theories have been displaced by others,
as the mean field \cite{Parisi1979,Parisi1983} and the droplet \cite{Fisher1986} pictures.
Partly because latter theoretical frameworks allow to derive analytical expressions 
for different quantities characterizing the physical behavior of a spin glass which, 
in principle, could be verified by comparing with experimental and numerical evidence.
Also, because the experience gained from numerical simulation indicates 
that these clusters, built on the basis of a ``local energy criterion'',
are not adequate to describe such disordered and frustrated systems.  
Nevertheless, despite the great theoretical progress made so far, the controversy 
about the nature of the spin-glass phase remains unresolved. 

In this context, it is not surprising that the efforts to describe 
the spin-glass problem in terms of magnetic clusters still persist.  
They are encouraged both by the experimental data, and the observation of 
strong spatial heterogeneities in numerical simulations of model systems. 
Such is the case of the Edwards-Anderson model \cite{EA} with bimodal couplings \cite{Toulouse1977}.
The ground state of this system is highly degenerated enabling to 
separate the lattice into two regions, one that contributes to the degeneration
and other that remains unchanged in all ground states.  The latter,
known as the ``rigid lattice''\cite{Barahona1982}, is a good candidate 
for a magnetic backbone capable of sustaining order in the system and has been the subject of various topological studies~\cite{Barahona1982,Klotz1994,Valdes1998,Vogel1999,Ramirez-Pastor2000}.   
In recent years, our understanding of the role played by this structure 
has advanced considerably.  In fact, for the case of the Edwards-Anderson model with bimodal couplings, spatial heterogeneities characterized by the rigid lattice have been associated to the equilibrium and 
nonequilibrium response of the system~\cite{Roma2006,Roma2007a,Roma2007b,Roma2010a,Roma2010b,Rubio2010a,Rubio2010b}.        
     
These studies have resulted in a generalization of the cluster approach for spin glasses (and for other systems with quenched disorder), appropriately named the ``backbone picture'' \cite{Roma2010b,Roma2013}.
The main conjecture of this picture is that it is possible to define an ``effective interaction'' between a pair of spins at sites $i$ and $j$, different from the bond strength $J_{ij}$. This quantity is the rigidity $r_{ij}$, which accounts for the energetic cost to leave the bond's state in the ground state (satisfied or frustrated).

The rigid structure is defined as the lattice where each bond $J_{ij}$ is replaced by its corresponding rigidity  $r_{ij}$. This can be accessed for spin-glass models with Ising spin variables~\cite{Roma2010b,Roma2013}, 
based on the ground-state configurations and the low-excitation levels.
In particular, in Ref.~\cite{Roma2013} the topological properties of the rigid structure were studied for the Edwards-Anderson model with both bimodal and Gaussian bond distributions, in two and three-dimensions. 
There, it was also analyzed how the spatial heterogeneities, characterized 
by the rigid structure, influence the equilibrium properties of these models
at finite temperature.  The results show that any disordered sample 
can be divided into two sectors, the backbone and its complement, 
and that within these regions the physical behaviors are very different.
What remains to be studied is how the rigid structure influences the nonequilibrium dynamics of the Edwards-Anderson model with Gaussian couplings.  
In order to fill this gap, in this work we study by means of
Monte Carlo simulations, the nonequilibrium dynamical heterogeneities 
of the three-dimensional Edwards-Anderson spin-glass model with Gaussian couplings within the framework given by the backbone picture.
Different two-times quantities, as the flipping time distribution and the 
correlation and response functions, are analyzed.  We find that the
dynamical heterogeneities can be clearly associated to the rigid structure.

The paper is structured as follows. In Sec.~\ref{ModRig} we present the 
Edwards-Anderson spin-glass model. We also summarize the most important properties
of its rigid structure and we introduce the concept of the backbone.  
Then, an extensive numerical simulation study of the nonequilibrium dynamics of this system 
is presented in Sec.~\ref{NumRes}. It is shown that different observables as 
the mean flipping time distribution, the correlation and the integrated response functions, 
as well as the fluctuation-dissipation plots, behave very differently
if they are evaluated inside or outside the backbone.
In the last section we discuss our results and conclusions are drawn. 

%.....................................................................
\section{The Edwards-Anderson spin-glass model and its ground state rigid structure \label{ModRig}}

We start by considering the three-dimensional (3D) Edwards-Anderson spin-glass model \cite{EA}
with Gaussian couplings (EAG), which consists of a set of $N$ Ising spins 
$\sigma_i = \pm 1$ placed in a cubic lattice of linear dimension $L$ ($N=L^3$), 
with periodic boundary conditions in all directions. The Hamiltonian is
\begin{equation}
H = - \sum_{(i,j)} J_{ij} \sigma_{i} \sigma_{j},  \label{Ham}
\end{equation}
where $(i,j)$ indicates a sum over the six nearest neighbors. 
The coupling constants or bonds, $J_{ij}$, are independent random variables 
drawn from a Gaussian distribution with mean value zero and variance one.  
The EAG model undergoes a continuous phase transition at a critical temperature 
$T_c \approx 0.95$ \cite{Katzgraber2006,Roma2010c} (temperatures 
are given in units of $1/k_B$, where $k_B$ is Boltzmann's constant).

Since the rigid structure and the backbone play a central role in this work,
now we give a brief summary of how these concepts have evolved in the last few years.
A more detailed discussion can be found in Refs.~\cite{Roma2010b,Roma2013}.  

For the simpler version of the model, the Edwards-Anderson spin-glass 
model with bimodal $\pm J$ couplings (EAB) \cite{Toulouse1977}, 
the ground state (GS) is multiply degenerated and then it is trivial to define a backbone. 
Bonds which do not change its condition (satisfied or frustrated) in all the 
configurations of the GS, are called ``rigid bonds'' and form the 
``rigid lattice'' \cite{Barahona1982}.
This structure of bonds and the set of spins connected by them constitute the so called backbone of the system.
The remaining bonds are called ``flexible bonds'' and form the ``flexible lattice''.

The backbone of the EAB model has some remarkable topological properties
that have been carefully evaluated through numerical simulations.
For two-dimensional (2D) square lattices, numerical results
are affected by significant finite size effects
but show that the most likely scenario is that
this structure does not percolate in the thermodynamic limit
(another possibility is that the backbone is just located on the percolation threshold).  
This backbone is fragmented into several clusters whose size distribution 
exhibits a power-law dependence, and the estimated critical exponents are 
consistent with the 2D random percolation universality class \cite{Stauffer1985}.
On the other hand, for the 3D cubic lattice both sets, the backbone and
its complement, percolate the system but with critical exponents differing from those within
the 3D random percolation universality class.  

More interesting is the connection between the backbone structure
and the nonequilibrium dynamical behavior.  
Both, the 2D and the 3D EAB models have a very broad distribution of relaxation times which, 
at low temperatures, show a spontaneous time-scale separation enabling to split the system
in two sets, one consisting of ``fast'' spins and another of ``slow'' spins \cite{Ricci2000}.
It was shown that these fast and slow degrees of freedom are directly 
related to spins outside and inside the backbone structure, respectively \cite{Roma2006,Roma2010a}. 
Furthermore, numerical simulations \cite{Barrat98,Ricci2003} 
show that below its critical temperature the 3D EAB model
exhibits a violation of the fluctuation-dissipation theorem (FDT) that, 
in the framework of mean field theory, resembles 
the full replica-symmetry breaking solution \cite{Cugliandolo2002,Crisanti2003}.   
We had previously shown that for very long simulation times spins outside the backbone
satisfy the FDT, whereas those within the backbone violate this relation \cite{Roma2007b}.  
Therefore, from a topological and dynamical point of view and below the critical temperature, a spin glass can be thought of as consisting 
of two different sets, one associated with an ordered-like phase (the backbone) and
another with a disordered paramagnetic-like phase (the complement of backbone).
Such results for the EAB model are not trivial.
In addition, a similar phenomenon is probably taking place in other spin-glass systems such
as the 2D $\pm J$ Potts model \cite{Ferrero2012} and 
the Viana-Bray model \cite{Montanari2003} as suggested by numerical results.

Notice that the definition of backbone based on the rigid lattice as used for the EAB model,
is not appropriate for systems with a simply degenerated GS,
since then the backbone structure would include the entire lattice. 
Such is the case of the EAG model which has a zero-entropy fundamental level 
(with only two configurations related by a global spin-flip) \cite{Perez2012}. 
In order to address the case of Ising models with simple degenerated GSs, 
a generalization of the concept of rigidity was presented in Refs.~\cite{Roma2010b,Roma2013}.
The definition of backbone is then based on the following concepts.  
The first key idea is to consider not only the GS but also the low-excitation levels.  
The ``rigidity'' of each bond should be a parameter taking a continuum of values, 
instead of only two (rigid-flexible) as in the EAB model.  
Let us consider a sample (a particular realizations of bond disorder) 
of the Edwards-Anderson model with an arbitrary bond distribution (discrete or continuous).  
For each bond $J_{ij}$ we define its rigidity as $r_{ij}=U_{ij}-U$, 
where $U$ is the GS energy of the sample and $U_{ij}$ is 
the lowest energy for which the bond $J_{ij}$ changes its GS condition 
(from satisfied to frustrated or vice versa). 
In other words, such bond does not change (remains rigid) 
for all configurations with energy lower than $r_{ij}+U$.  
The ``rigid structure'' (RS) of a sample is then defined
as the lattice where each bond $J_{ij}$ has been replaced by its rigidity $r_{ij}$.  
According to the backbone picture, the latter 
is a quantity that gives the magnitude of the ``effective interaction''  
between the pair of spins at sites $i$ and $j$.  

\begin{figure}[t!]
\begin{center}
\includegraphics[width=7.5cm,clip=true]{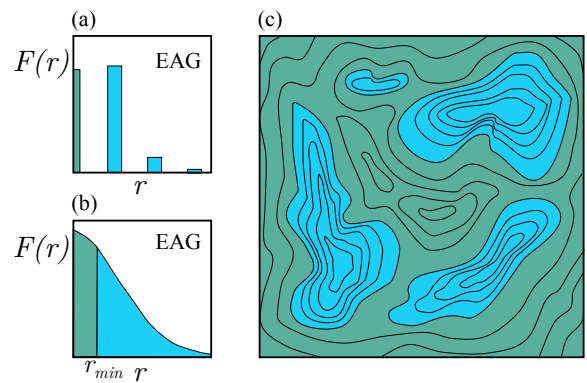}
\caption{\label{fig1} Rigidity distribution for the (a) EAB and (b) EAG models. 
Different colors indicate rigidities below and above the threshold value, $r_{\mathrm{min}}$. 
(c) Two dimensional schematic picture of the spatial distribution of rigidity values. 
Lines correspond to contour levels while the colors indicate high and low rigidity 
regions according to the rigidity distributions shown in (a) and (b).}
\end{center}
\end{figure}

The RS for both the EAB and the EAG models, can be numerically determined 
using a method proposed in Refs.~\cite{Roma2010b,Ramirez2004}. 
For this approach to work properly, it is necessary to find
many low-lying energy configurations for each sample. 
For the EAG model the number of configurations one needs 
to determine is of the order of the number of bonds in the system.  
In Ref.~\cite{Roma2013}, a parallel tempering Monte Carlo algorithm \cite{Geyer1991,Hukushima1996} 
was used for this purpose.  
This technique, widely used in equilibrium simulations,
also provides a powerful heuristic method for reaching the GS 
of spin-glass models \cite{Moreno2003,Roma2009}.  
Nevertheless, because the calculation of GS configurations 
of the 3D Edwards-Anderson model is an NP-hard problem, 
in practice the RS can be numerically determined only for samples of small size.  
In particular, for the 3D EAG model, the parallel tempering algorithm allows one
to obtain GSs with a high probability for lattice sizes up to $L=10$ \cite{Roma2009}.
Since $3N$ of such configurations are required to calculate the RS of a single sample,
we restrict our analysis to $L=8$ \cite{Roma2013}.

The interpretation of the rigidity $r_{ij}$ as an effective interaction
is the basis on which rests a generalized definition of the backbone.
Given the random distribution of bonds, the rigidity distribution $F(r)$, 
where $F(r) dr$ is the fraction of bonds whose rigidities are between $r$ and $r+dr$, can be computed.
The results, taken from Ref.~\cite{Roma2013}, are shown in 
Figs.~\ref{fig1}(a) and (b) for the 3D EAB and EAG models, respectively.
For the EAB model the rigidity distribution is discrete and bonds 
with non-zero rigidity [$r=4$, $r=8$, and $r=12$, indicated with a different color in Fig.~\ref{fig1}(a)] 
have similar physical behaviors. Thus, naturally, 
they are considered as parts of a single set, determining the main physical behavior, which we call the backbone.  
The remaining bonds with $r=0$, behave in a very different way 
and are considered outside this set. 

For the EAG model the rigidity is a continuous variable and 
the rigidity distribution takes the continuum form presented in Fig.~\ref{fig1}(b). 
Since we expect that the physical behavior of the 3D Edwards-Anderson model should
not depend on the exact form of the bond distribution,
then even for this continuous rigidity distribution 
it should be possible to divide the system into two parts with different physical properties: the backbone and its complement.    
In fact, as shown in Ref.~\cite{Roma2013}, in 2D a rigidity threshold value $r_{\mathrm{min}}$ can be found
to separate the system into high-rigidity (HR) and low-rigidity (LR) components
such that the EAB and the EAG models
share the same percolation properties and have a similar equilibrium behavior.
For example, a $r_{\mathrm{min}}$ value can be selected such that a similar temperature dependence 
of the average energy per bond within each region for the EAB and the EAG models can be obtained.
Therefore, bonds within the HR region (which has $r \ge r_{\mathrm{min}}$) form the backbone of the system.
Furthermore, this idea can be extended to 3D, where following these same criteria 
the numerical data suggest that a suitable backbone could be defined
for $0.6 < r_{\mathrm{min}} < 2.0$.
Based on these results, and also to minimize frustration within the HR region,
we use here the value $r_{\mathrm{min}}=1.3$ (which was already used in Ref.~\cite{Roma2013}).
Moreover, it is important to highlight that 
by choosing values of $r_{\mathrm{min}}$ in the range $0.6-2.0$ we obtain
qualitatively similar results (as discussed in the next Section and in the Appendix),
thus not changing our main findings.

We stress that although bond values are randomly 
distributed on the sample, rigidity values are spatially correlated. 
In Fig.~\ref{fig1}(c) the spatial distribution of rigidity values 
are schematically shown using contour levels and with colors indicating HR and LR regions. 
Note that this figure is a schematic representation introduced with the intention 
of presenting the main ideas leading to a backbone picture
(the map plots in figure 2 of Ref.~\cite{Roma2013}, show actual images
of the spatial distribution of rigidity in 2D).  
In the framework of this phenomenological theory, we emphasize that 
bonds and spins within these regions
should display different dynamical behaviors.

In the following section, we study the strong heterogeneities arising in
the nonequilibrium dynamics of the 3D EAG model, 
by evaluating different spin observables over the full (all spins), the HR 
(spins connected by bonds with $r \ge r_{\mathrm{min}}=1.3$), and the 
LR (remaining spins) regions.  To this end, we use the same set of
$10^3$ samples of $L=8$ considered in Ref. \cite{Roma2013}, 
for which the corresponding RSs have already been calculated. 

%.....................................................................
\section{Numerical results \label{NumRes}}

In order to investigate the nonequilibrium dynamics of the 3D EAG model, 
we use a typical protocol which consists on a quench at time 
$t=0$ from a random configuration to a low temperature $T$ below $T_c$. 
From this initial condition, the evolution of the system is simulated 
by a single spin-flip Monte Carlo algorithm
and different two-times quantities are analyzed, which depend on 
both the waiting time $t_{\mathrm{w}}$, when the measurement begins, 
and a given time $t>t_{\mathrm{w}}$.
A unit of time consists of $N$ elementary random spin-flip attempts. 
As stated above, in all cases cubic lattices of linear size 
$L=8$ with full periodic boundary conditions were simulated.
In addition, in some cases simulations on larger lattices of $L=20$ 
were carried out in order to address finite size effects.
The disorder average was performed over $10^3$ different samples.
Furthermore, for each sample we have averaged over 
$10$ independent thermal histories (or runs), i.e.,
along different initial conditions and realizations of the thermal noise
(except in subsection \ref{secFDT}, where $10^4$ of such runs were required
to calculate the FDT map plots).

We simulate each sample for six temperatures between $T=0.4$ and $T=0.9$,
for different waiting times up to $t_{\mathrm{w}}=10^7$, and for a
maximum time of $t=2 \times 10^7$.  
Even for the highest temperature and within the longest time window, 
all observables calculated in this study depend on $t_{\mathrm{w}}$,
indicating that our simulations are far from the equilibrium regime.

Hereinafter, we analyze the correlation and response functions, 
as well as the corresponding FDT plots. 
Then, the flipping time distribution for different temperatures 
and waiting times is studied.  

\subsection{Correlation and response functions}

We begin by calculating the full two-times correlation function defined for $t>t_{\mathrm{w}}$ as
\begin{equation}
C (t,t_{\mathrm{w}}) = \left [ \left \langle \frac{1}{N} \sum_{i=1}^N
\sigma_i(t) \sigma_i(t_{\mathrm{w}}) \right \rangle \right
]_{\mathrm{av}}, \label{corr}
\end{equation}
where $\langle \dots \rangle$ represent an average over thermal histories 
and $[\dots]_{\mathrm{av}}$ indicates average over disorder realizations. 
On the other hand, by adding to the Hamiltonian (\ref{Ham}) a perturbation of the form
$H_p=-\sum_{i=0}^N h_i \sigma_i$, 
with $h_i=\pm h$ being a random field of 
intensity $h$ switched on at time $t_{\mathrm{w}}$, 
it is possible to calculate the full two-times integrated response function as \cite{Crisanti2003}
\begin{equation}
\chi (t,t_{\mathrm{w}}) = \left [ \left \langle \frac{1}{Nh}  \sum_{i=1}^N \sigma_i(t) \
\mathrm{sgn}(h_i(t_{\mathrm{w}})) \right \rangle_h \right ]_{av}. \label{resp}
\end{equation}
Here, the average $\langle \dots \rangle_h$ should be taken
over thermal histories of the perturbed system, 
and the value of $h$ should be chosen small enough 
in order to reach the linear response regime.
Nevertheless, instead of performing additional simulations with an applied field, 
we calculate $\chi$ for infinitesimal perturbations 
using the algorithm proposed in Refs.~\cite{Ricci2003,Chatelain2003},
which permits to determine the correlation and the 
response functions in a single simulation of the unperturbed system.   
This algorithm needs a Monte Carlo simulation implemented with 
a differentiable transition rate of the external field.
Then, we use Glauber dynamics where local changes 
are accepted with a transition rate given by \cite{Glauber1963}
\begin{equation}
W_{\mathrm{G}} = \frac{\exp \left( - \beta \Delta H \right) }{1 + \exp
\left( - \beta \Delta H \right)}, \label{rate}
\end{equation}
where $\beta=1/T$ and $\Delta H$ is the energy change required to flip a given spin.

\begin{figure}[t!]
\begin{center}
\includegraphics[width=7.3cm,clip=true]{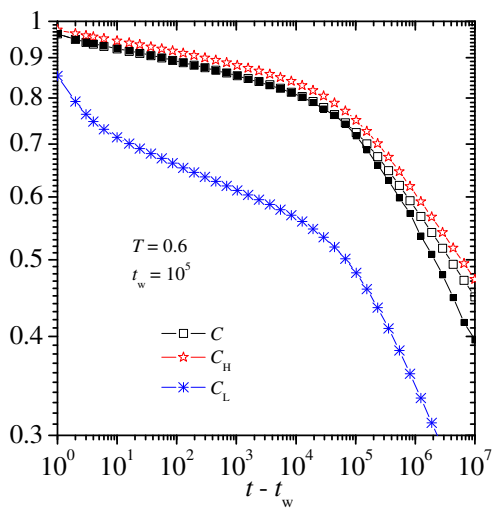}
\caption{\label{fig2} The correlation functions 
$C$, $C_{\mathrm{H}}$, and $C_{\mathrm{L}}$ at $T=0.6$ for the waiting time $t_{\mathrm{w}}=10^5$.
Closed black squares correspond to the full correlation function calculated for 
samples of size $L=20$.}
\end{center}
\end{figure}

Figure~\ref{fig2} shows the dependence of the full correlation function $C$
with $t-t_{\mathrm{w}}$, for $t_{\mathrm{w}}=10^5$ at temperature $T=0.6 < T_c$.
Open (closed) black squares correspond to simulations for systems of
size $L=8$ ($L=20$). From this comparison we conclude that 
finite-size effects are not important and then
samples with $L=8$, for which we know their RSs, 
represent very well the dynamic behavior of the system. 
This is true for all quantities calculated in the following.  

Most interesting is the heterogeneous behavior of this quantity.
As we mentioned in the previous Section, each sample of the EAG system 
can be divided into the HR and LR regions. 
In addition to $C$, we then calculate the correlation function for
the HR and the LR regions, $C_{\mathrm{H}}$ and $C_{\mathrm{L}}$ respectively,
by restricting the sum in Eq.~(\ref{corr}) to those spins belonging to each sector
(and we normalize both sums using the corresponding number of spins of each region).  
Figure~\ref{fig2} shows that the behaviors of $C$ and $C_{\mathrm{H}}$ are very similar, 
while $C_{\mathrm{L}}$ displays a more drastic initial relaxation followed 
by a second decay process as fast as the full correlation function.  

\begin{figure}[t!]
\begin{center}
\includegraphics[width=7.3cm,clip=true]{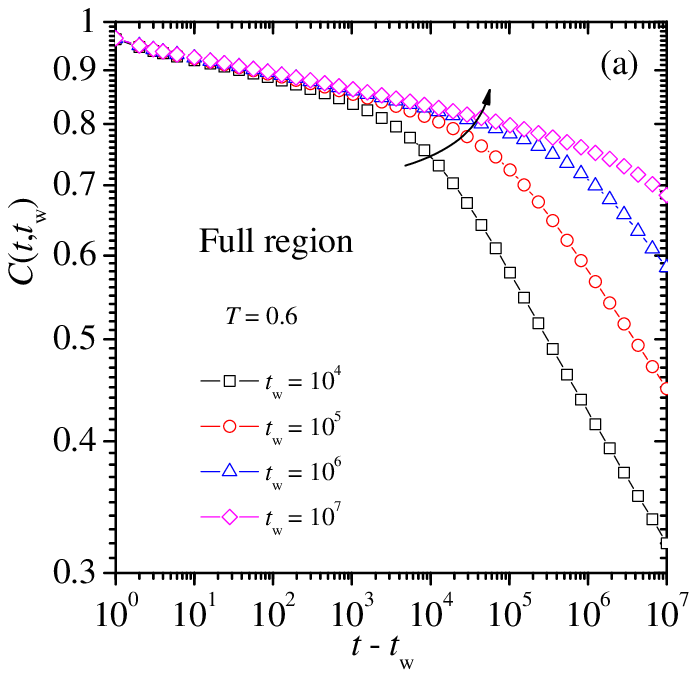}
\includegraphics[width=7.3cm,clip=true]{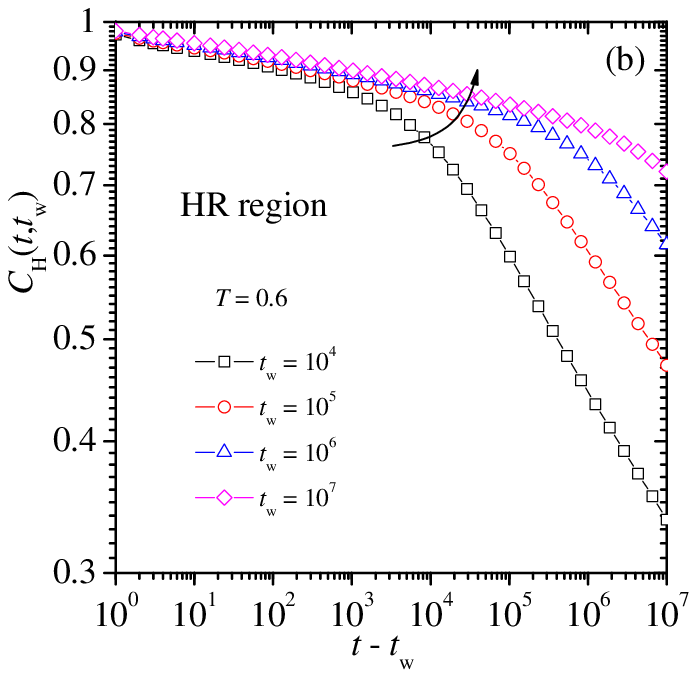}
\includegraphics[width=7.3cm,clip=true]{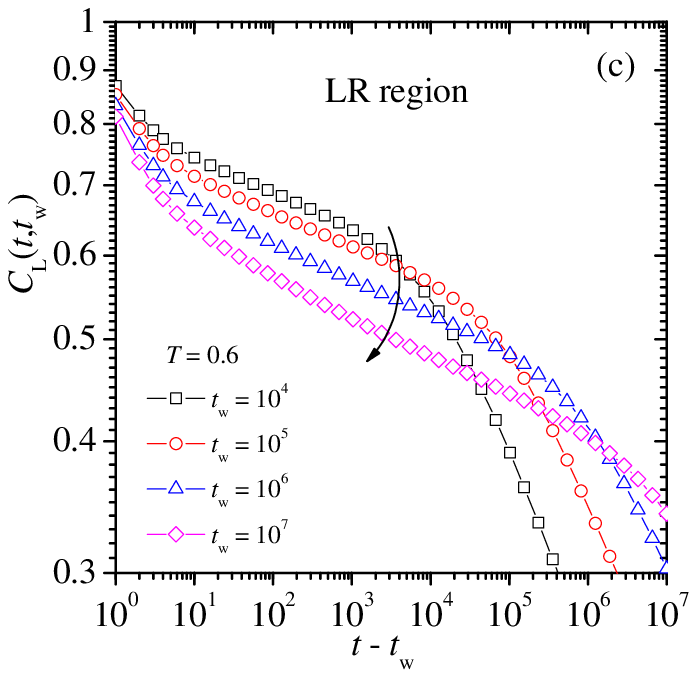}
\caption{\label{fig3} The correlation functions
(a) $C$, (b) $C_{\mathrm{H}}$, and (c) $C_{\mathrm{L}}$ at $T=0.6$ and 
for different waiting times as indicated.  
The arrows emphasize how such functions evolve with increasing $t_{\mathrm{w}}$.}
\end{center}
\end{figure}

Now, we analyze the dependence of correlation functions with the waiting time. Figure~\ref{fig3} (a) shows the full correlation function at $T=0.6$. As expected, this quantity displays a typical aging behavior.
Initially the correlation function slowly relaxes but, after a certain period of time, a faster relaxation process develops. With increasing $t_{\mathrm{w}}$ this second relaxation process begins at longer $t-t_{\mathrm{w}}$.  

In Figs.~\ref{fig3} (b) and (c) we show the plots of $C_{\mathrm{H}}$ and $C_{\mathrm{L}}$ at temperature $T=0.6$.
While $C_{\mathrm{H}}$ shows the same evolution as $C$ [compare with Fig.~\ref{fig3} (a)],
with increasing waiting time the first decay of the correlation $C_{\mathrm{L}}$
becomes more pronounced, indicating that the spins 
within the LR region will be less correlated with time. 
This behavior is also observed in the nonequilibrium dynamics of
3D EAB model below its critical temperature \cite{Roma2010a},
and indicates a significant difference between the backbone and its complement.
As we mentioned in Section~\ref{ModRig} and in the Appendix,
by choosing values of $r_{\mathrm{min}}$ in $0.6 < r_{\mathrm{min}} < 2.0$ we obtain similar results.

\begin{figure}[t!]
\begin{center}
\includegraphics[width=7.3cm,clip=true]{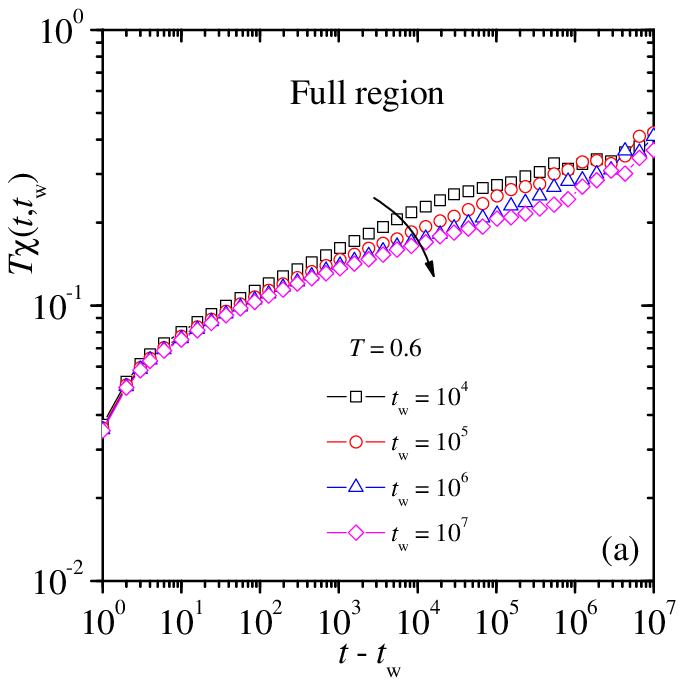}
\includegraphics[width=7.3cm,clip=true]{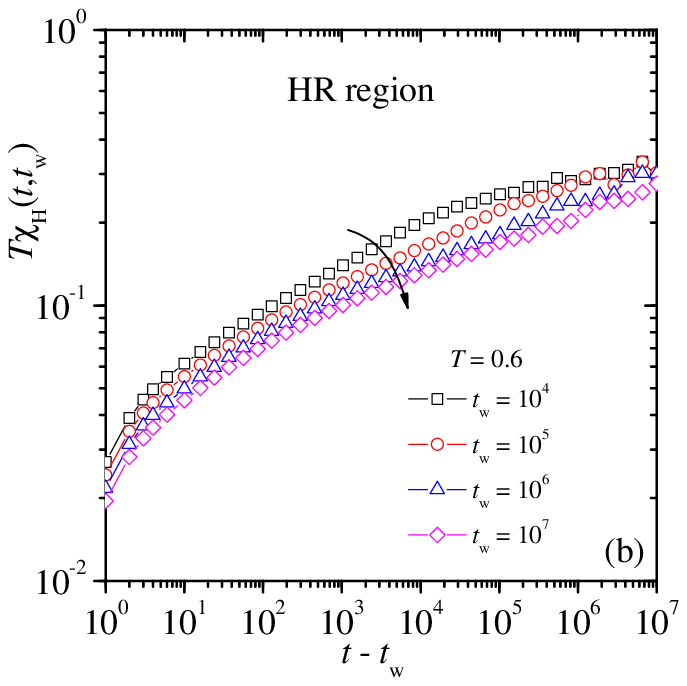}
\includegraphics[width=7.3cm,clip=true]{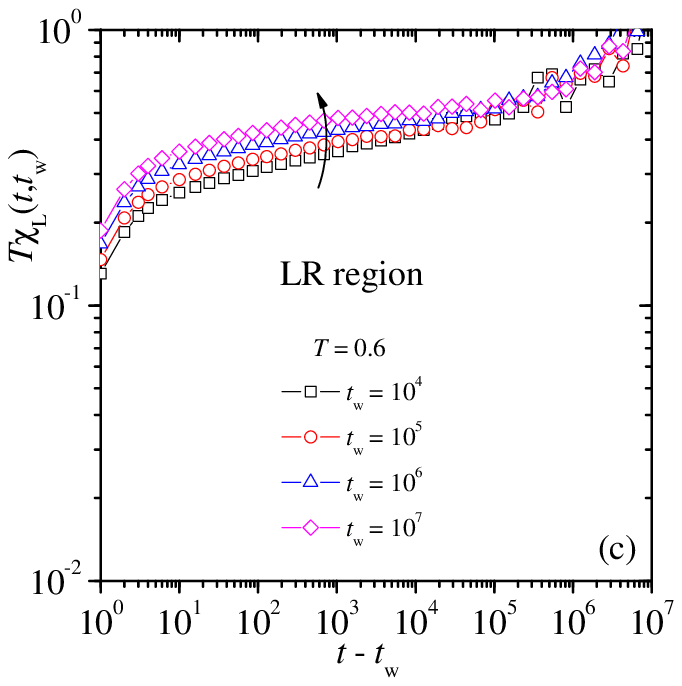}
\caption{\label{fig4} The integrated response functions 
(a) $\chi$, (b) $\chi_{\mathrm{H}}$, and (b) $\chi_{\mathrm{L}}$.
Curves are for $T=0.6$ and different waiting times as indicated.  
The arrows emphasize how such functions evolve with increasing $t_{\mathrm{w}}$.  }
\end{center}
\end{figure}

The full integrated response function $\chi$, as well as
$\chi_{\mathrm{H}}$ and $\chi_{\mathrm{L}}$ (the responses of spins 
belonging to the HR and the LR regions) 
are shown, respectively, in Figs.~\ref{fig4} (a), (b), and (c).
As before for the correlation functions calculated over the full and the HR regions, 
$\chi$ and $\chi_{\mathrm{H}}$ show a similar trend although in this case  
both fall instead of rising with increasing $t_{\mathrm{w}}$.
This is typical of nonequilibrium systems which develop
a sort of domain-growth process (for example coarsening in ferromagnets), 
where the domains poorly respond to an external perturbation 
while the walls that separate them are composed of spins that easily align with the applied field~\cite{Roma2010a}.
As the system evolves these domain walls tend to disappear and 
therefore the response function must decrease for long times. 
Again our numerical results reinforce the idea that a domain-growth process could be taking place inside the backbone. 

On the other hand, for short times the response of spins within the LR region
is larger than for the HR region, see Fig.~\ref{fig4} (c).
Besides, $\chi_{\mathrm{L}}$ increases with increasing $t_{\mathrm{w}}$, contrary to what happens for $\chi$ and $\chi_{\mathrm{H}}$.
This behavior shows that the LR region become disordered
and probably is not able to support an ordered phase.
In the next section, the analysis of the combined effect of 
the correlation and the integrated response functions, the FDT plots, 
will be crucial to understand the overall physical behavior of each region.       
  
\subsection{FDT plots \label{secFDT}}

The full correlation and integrated response functions
are related through a quasi-fluctuation-dissipation theorem  
\begin{equation}
T \chi(t,t_{\mathrm{w}})=X\left[ 1-C(t,t_{\mathrm{w}}) \right], 
\end{equation}
where $X$ is the fluctuation-dissipation ratio (FDR) \cite{Peliti1997}. 
A useful representation of this equation is a parametric plot, 
or ``FDT plot'', of $T\chi$ vs. $C$ \cite{Cugliandolo2002,Crisanti2003}. 
At thermodynamic equilibrium when the FDT holds, 
the FDR is $X=1$ and the parametric plot shows a
linear relation with a slope of $-1$. 
In this case the correlation and the 
integrated response functions depend on $\Delta t=t-t_{\mathrm{w}}$.
On the other hand, in a nonequilibrium situation the FDT does not longer hold 
and two regimes are observed: for $t/t_{\mathrm{w}} \ll 1$ the system shows quasi-equilibrium
with $X=1$, while for $t/t_{\mathrm{w}} \gg 1$ a violation of the FDT is
observed with $X<1$. The behavior of $X$ for $t/t_{\mathrm{w}} \gg 1$
allows for a simple classification of nonequilibrium systems
into three main categories: (\textit{i}) the value $X=0$ is
related to coarsening systems, (\textit{ii}) a constant $X<1$
value is associated with structural glasses, and (\textit{iii}) a
decreasing monotonic $0<X<1$ function corresponds to spin glasses. 

\begin{figure}[t!]
\begin{center}
\includegraphics[width=7.3cm,clip=true]{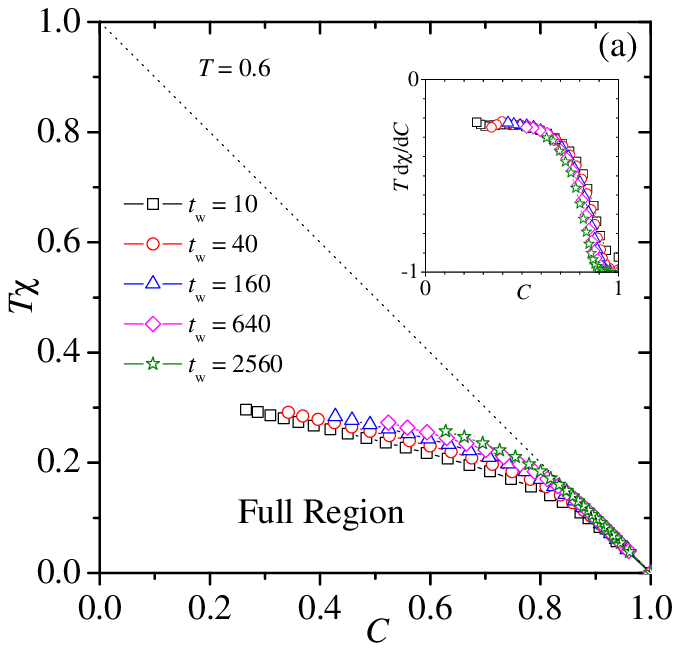}
\includegraphics[width=7.3cm,clip=true]{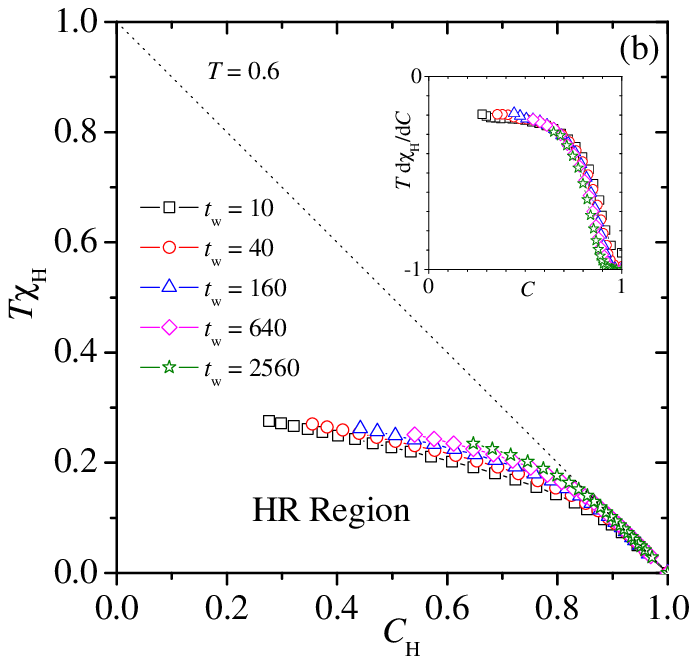}
\includegraphics[width=7.3cm,clip=true]{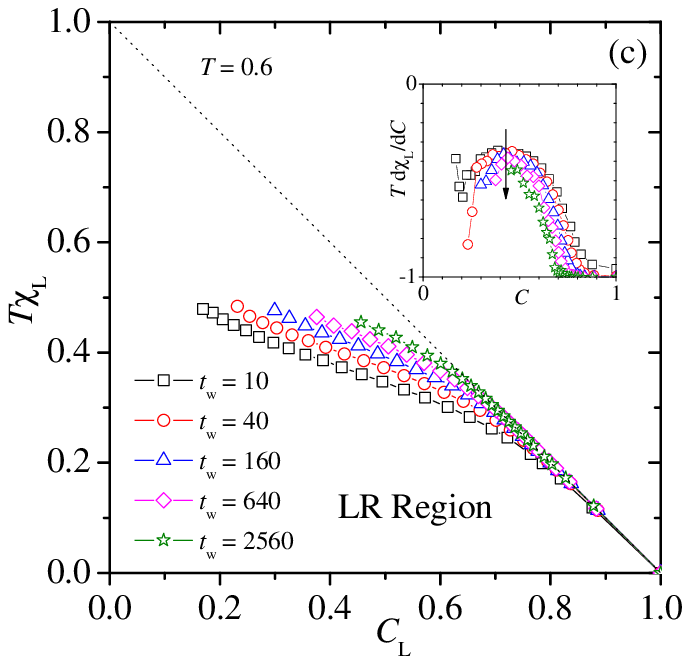}
\caption{\label{fig5} The FDT plots for (a) the Full, 
(b) the HR, and (c) the LR regions at $T=0.6$ and for different $t_{\mathrm{w}}$ as indicated. 
The insets show the slope of these curves as function of $C$.}
\end{center}
\end{figure}

In Fig.~\ref{fig5} (a) we can see the FDT plot for $T=0.6$ and different waiting times.
After the quasi-equilibrium regime, the curves deviates from the straight line of slope $-1$ 
and show a continuous variation of the FDR (see their derivatives in the inset)
which, as noted above, is associated to spin glasses \cite{Barrat98,Ricci2003}.
On the other hand, Fig.~\ref{fig5} (b) shows the FDT plot of the spins belonging to the HR region
for different values of $t_{\mathrm{w}}$.  Notice the similarity with the curves in Fig.~\ref{fig5} (a).  
There does not seem to be differences between the full system and the HR region.  

Nevertheless, a rather more striking behavior is observed for
the FDT plot within the LR region, Fig.~\ref{fig5} (c).
We expect this set to behave like the spins outside the backbone in the 3D EAB model, i.e.
fulfill FDT ($X \to 1$) for long times.
In fact, as shown in the inset of this figure,
the slope of these FDT curves tend to $-1$ for increasing $t_{\mathrm{w}}$.

\begin{figure}[t!]
\begin{center}
\includegraphics[width=8cm,clip=true]{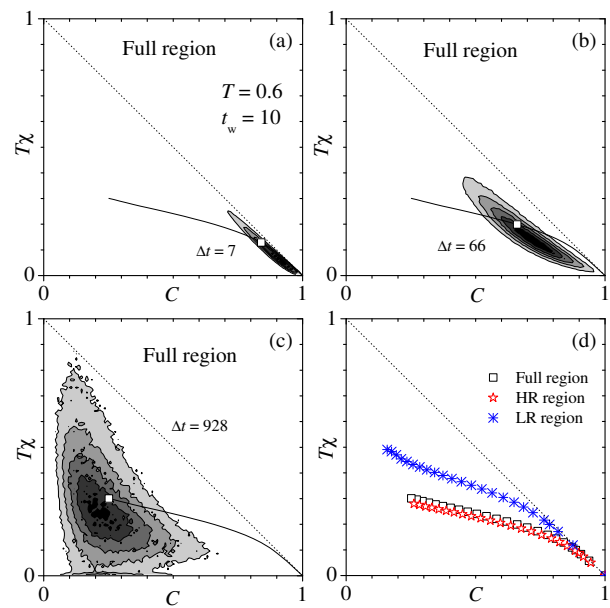}
\caption{\label{fig6} Panels (a), (b), and (c) show 
the map plots of the JPD for, respectively, $\Delta t=7$, $66$, and $928$,
together with the corresponding FDT plots (solid curves). 
Dark (bright) areas denote high (low) density points.   
The open squares indicates the position, for each $\Delta t$, of the first moment of the JPD.  
Panel (d) shows the FDT plots for the full, the HR, and the LR regions.    }
\end{center}
\end{figure}

In order to further analyze how the heterogeneities arise in the violation
of the FDT, we follow the lines of Ref.~\cite{Roma2007b}.
First, we note that the full correlation function (\ref{corr})
can be rewritten as
\begin{equation}
C (t,t_{\mathrm{w}}) = \left [ \frac{1}{N} \sum_{i=1}^N \widetilde{C}_i(t,t_{\mathrm{w}}) \right]_{\mathrm{av}}, \label{corr2}
\end{equation}
where 
\begin{equation}
\widetilde{C}_i(t,t_{\mathrm{w}}) = \left \langle \sigma_i(t) \sigma_i(t_{\mathrm{w}}) \right \rangle  \label{single_corr}
\end{equation}
is the thermal average of a single spin correlation function.
The same is true for the integrated response function (\ref{resp}), 
\begin{equation}
\chi (t,t_{\mathrm{w}}) =  \left [ \frac{1}{N} \sum_{i=1}^N \widetilde{\chi}_i(t,t_{\mathrm{w}}) \right]_{\mathrm{av}}. \label{single_resp}
\end{equation}
Then, the sample average of the joint probability distribution (JPD),  
$\rho(\widetilde{C},T \widetilde{\chi},t,t_{\mathrm{w}})$,
allows us to visualize clearly how each spin contributes to $C$ and $\chi$.
The full correlation and integrated response functions can be recovered by simple integration,
\begin{equation}
C (t,t_{\mathrm{w}}) = \int d\widetilde{C} \int d\widetilde{\chi} \
\rho(\widetilde{C},T \widetilde{\chi},t,t_{\mathrm{w}}) \widetilde{C}  \label{corr3}
\end{equation}
and  
\begin{equation}
\chi(t,t_{\mathrm{w}}) = \int d\widetilde{C} \int d\widetilde{\chi} \
\rho(\widetilde{C},T \widetilde{\chi},t,t_{\mathrm{w}}) \widetilde{\chi} . \label{res3}
\end{equation}

Figures~\ref{fig6} (a)-(c) show different map plots of the JPD at $T=0.6$ and for $t_{\mathrm{w}}=10$.
The sequence represents the evolution of this distribution for increasing $\Delta t$, 
together with the corresponding FDT plots (solid curves).
Also, for each case, the corresponding coordinates 
of the first moment of the JPD given by Eqs.~(\ref{corr3}) and (\ref{res3}) are shown. 
We emphasize that these map plots were calculated for $10^3$ samples and, in order
to obtain a converged distribution it was necessary to carry out $10^4$ independent runs 
\footnote{The product $\sigma_i(t) \sigma_i(t_{\mathrm{w}})$ can take only two values, $\pm 1$. 
But the average of this quantity, the single spin correlation function (\ref{single_corr}), 
has a real value with a standard deviation which is sufficiently small ($\ll 1$),
if an adequate number of independent runs are used to calculate it. 
The same is true for the single response function.}.  

At very short $\Delta t$ into the quasi-equilibrium regime,
Fig.~\ref{fig6} (a) shows that the JPD is compact and 
has an elongated shape oriented along the straight line of slope $-1$. 
This behavior can be attributed exclusively both, 
to the pronounced fall of the correlation function 
and to the strong response to an external perturbation,
that is observed for short times within the LR region.
However, with increasing $\Delta t$ the distribution widens in all directions 
reflecting the increase of dynamical heterogeneities [see Figs.~\ref{fig6}(b) and (c)].
This broadening of the JPD can be interpreted analysing separately 
the behavior of the entire system, the backbone and its complement.     
In Fig.~\ref{fig6} (d) we can see the FDT plots for the full ($T\chi$ vs. $C$), 
the HR ($T\chi_{\mathrm{H}}$ vs. $C_{\mathrm{H}}$), 
and the LR ($T\chi_{\mathrm{L}}$ vs. $C_{\mathrm{L}}$) regions. 
The first two show a similar flattening trend, but the LR region shows a pronounced tendency towards equilibrium behavior.
This result reinforces the idea proposed in Ref.~\cite{Roma2007b}, that the complement of backbone remains in a paramagnetic phase even below the critical temperature.

\subsection{Flipping time distribution}

Finally, we shall focus on the mean flipping time distribution as an effective direct
way to analyze the strong time-scale separations observed in the EAG model. 
For each sample, we measure the number of flips, $N_{\mathrm{F}}$, 
done by every spin within a time window extending 
from $t_{\mathrm{w}}$ to $t$, being $\Delta t = t-t_{\mathrm{w}}$. 
The mean flipping  time $\tau_{\mathrm{F}}$ for a given $t_{\mathrm{w}}$ and $t$ 
is defined as the time window size divided by the number of flips: 
$\tau_{\mathrm{F}}(t,t_{\mathrm{w}})= \Delta t/ N_{\mathrm{F}}$ \cite{Ricci2000}. 
We calculate the mean flipping time distribution for each sample 
and then we average these to compute $P(\ln \tau_{\mathrm{F}})$.
As in Ref.~\cite{Ricci2000}, we use a logarithmic scale for the argument due to the broadness of this distribution.

Instead of Glauber, we choose to work with a Metropolis dynamics 
whose transition rate is given by \cite{Metropolis}
\begin{equation}
W_{\mathrm{M}}=\mathrm{min} \{1,\exp(-\beta \Delta H) \}  . \label{Metropolis}
\end{equation}
Although both dynamics belong to the same universality class (Model A) \cite{Hohenberg1977},
we prefer to use Metropolis to make a direct comparison 
with the results reported in Refs.~\cite{Ricci2000,Roma2006,Roma2010a} for the EAB model.

\begin{figure}[t!]
\begin{center}
\includegraphics[width=7.3cm,clip=true]{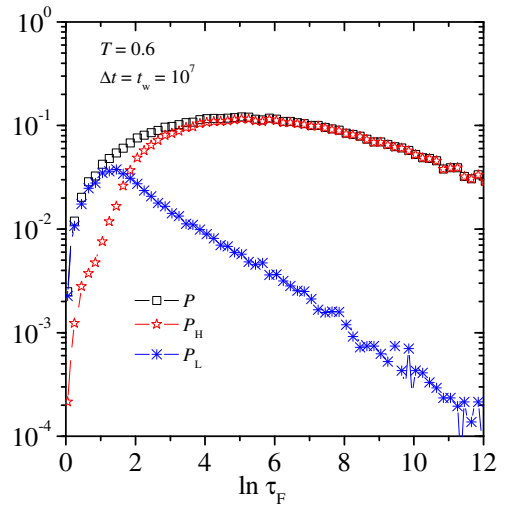}
\caption{\label{fig7} The mean flipping time distribution 
for $\Delta t=t_{\mathrm{w}}=10^7$ at $T=0.6$.  Figure shows a comparison 
between the full distribution $P$ and the corresponding ones for the HR 
and the LR regions, $P_{\mathrm{H}}$ and $P_{\mathrm{L}}$, respectively.}
\end{center}
\end{figure}

When analyzing the 3D EAB model, Ricci-Tersenghi and Zecchina showed that the 
flipping time distribution widens as the temperature decreases and develops a visible 
two-peaked structure \cite{Ricci2000}.  
As the temperature decreases, the peak located at high flipping times, 
corresponding to the slow degrees of freedom, moves towards higher 
values in accordance with an activation process of energy $\varepsilon = 4$ 
\footnote{For the bimodal Edwards-Anderson model with $\pm J$ couplings, 
the minimum excitation energy is $\varepsilon = 4J$; here we have taken $J=1$. }. 
Concurrently, the peak located at small flipping times, which characterizes fast degrees 
of freedom, does not depend on temperature and is located at $\tau_{\mathrm{F}} \approx 1$.  
This means that the EAB model is composed of a fraction of spins which behave like
approximately ``free spins'', i. e. spins whose flipping does not change 
the energy and then can be flipped on every time step 
(the Metropolis rate for accepting these moves is one). 
Within the backbone picture this set of spins plays an important role because,
even well below $T_c$, it behaves like a paramagnet. 

To study the 3D EAG model we compare, as before, the full mean flipping time distribution $P$
with the corresponding ones $P_{\mathrm{H}}$ and $P_{\mathrm{L}}$ calculated over the sets of spins belonging to the HR and the LR regions.  
Figure~\ref{fig7} shows these functions at $T=0.6$ and for $t_{\mathrm{w}}=\Delta t=10^7$.
Although a strong time-scale separation (two-peaked structure) is not observed,
we clearly obtain that the LR (HR) region represent very well the fast (slow) degrees of freedom, as in the EAB model

\begin{figure}[t!]
\begin{center}
\includegraphics[width=7.3cm,clip=true]{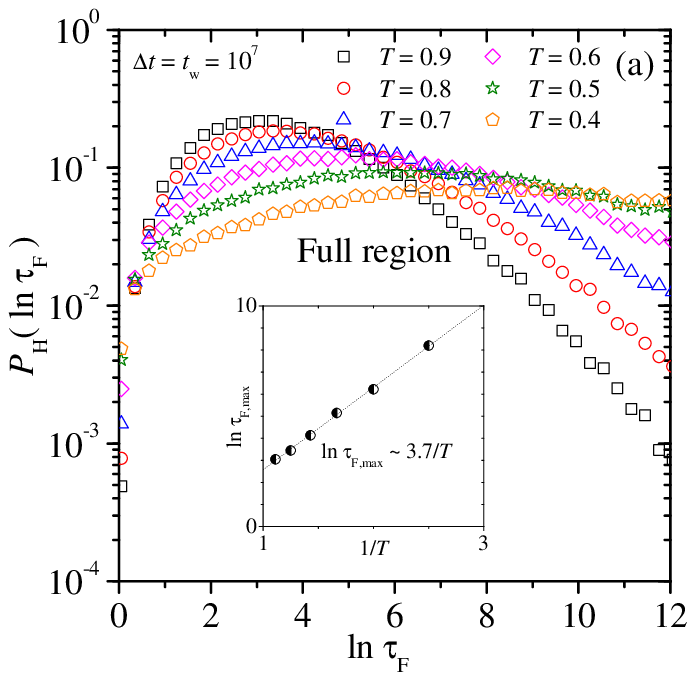}
\includegraphics[width=7.3cm,clip=true]{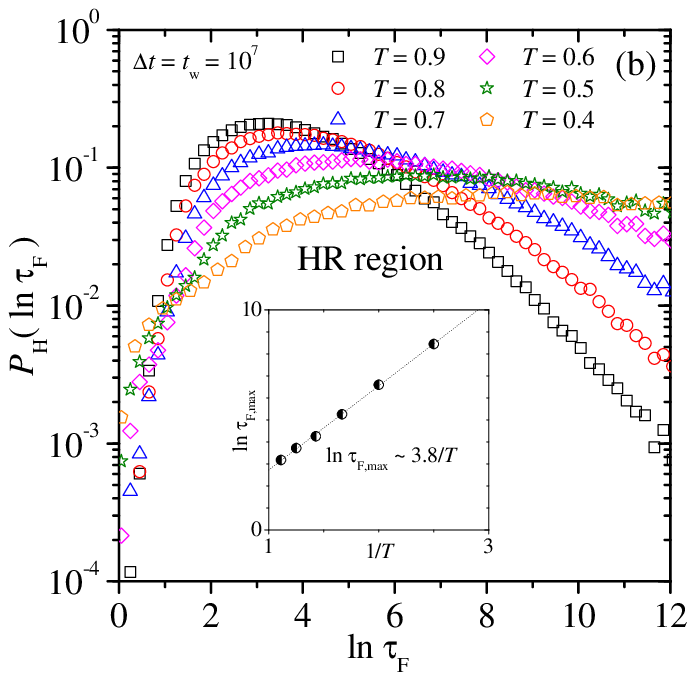}
\includegraphics[width=7.3cm,clip=true]{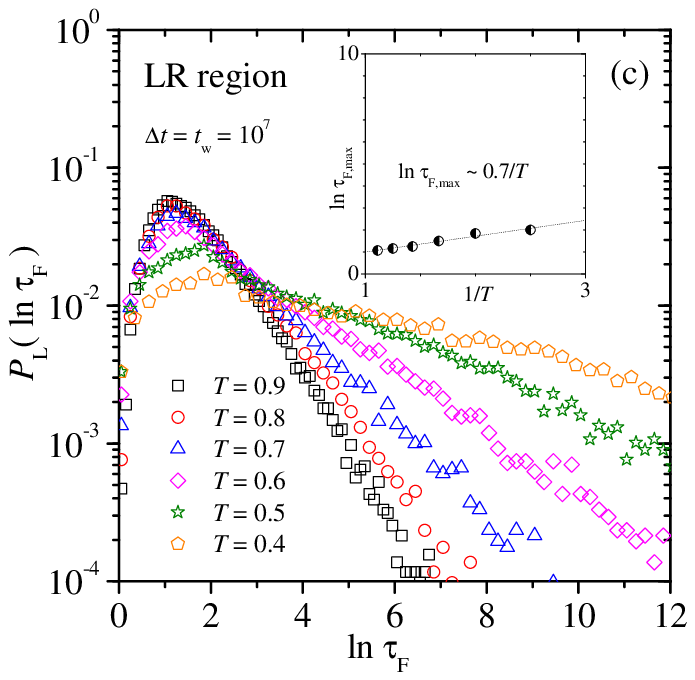}
\caption{\label{fig8} The mean flipping time distributions
(a) $P$, (b) $P_{\mathrm{H}}$, and (c) $P_{\mathrm{L}}$ for $\Delta t=t_{\mathrm{w}}=10^7$  
and for different temperatures as indicated.  The insets present the dependency of 
$\ln \tau_{\mathrm{F,max}}$ with $1/T$ for each region. }
\end{center}
\end{figure}

Figure~\ref{fig8}(a-c) show the mean flipping time distribution 
for different temperatures below the critical one, and again for $t_{\mathrm{w}}=\Delta t=10^7$.  
For the full set of spins, Fig.~\ref{fig8}(a), we observe that as temperature is decreased $P$ becomes broader and more spins begin to flip slowly. The same qualitative behavior is observed for $P_{\mathrm{H}}$ in Fig.~\ref{fig8}(b). This is a typical low-temperature behavior observed in several glassy systems.  
Note also that the single peaks of $P$ and $P_{\mathrm{H}}$ depend on temperature as $\ln \tau_{\mathrm{F,max}} \sim \varepsilon/T$, with $\varepsilon = 3.7$ and $3.8$, respectively (see insets).
This behavior is very close to the corresponding one for the slow 
degrees of freedom of the EAB model, for which the position of the peak for large flipping times scales as $4/T$ \cite{Ricci2000,Roma2006,Roma2010a}.  
On the other hand, for the LR region [Figs.~\ref{fig8} (c)] the peak of the distribution $P_{\mathrm{L}}$ becomes sharper and scales also as $\ln \tau_{\mathrm{F,max}} \sim \varepsilon/T$ but with a smaller value $\varepsilon = 0.7$.

\begin{figure}[t!]
\begin{center}
\includegraphics[width=7.3cm,clip=true]{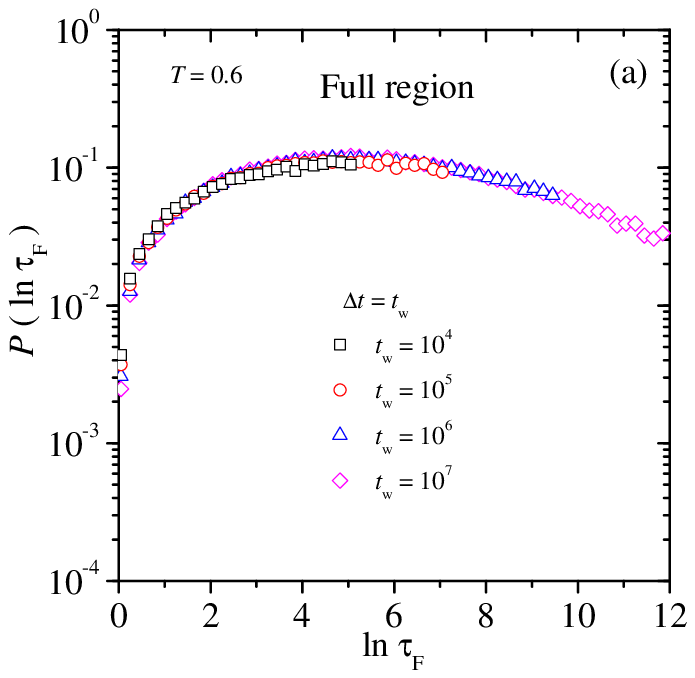}
\includegraphics[width=7.3cm,clip=true]{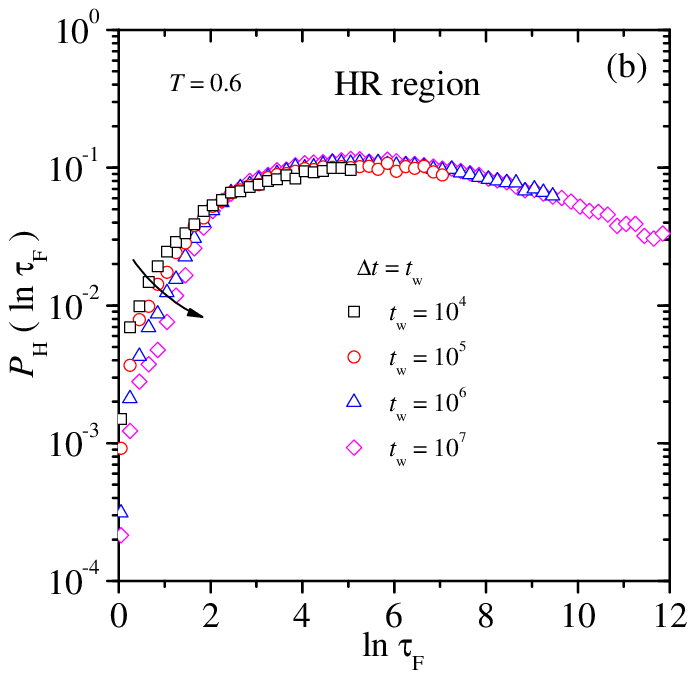}
\includegraphics[width=7.3cm,clip=true]{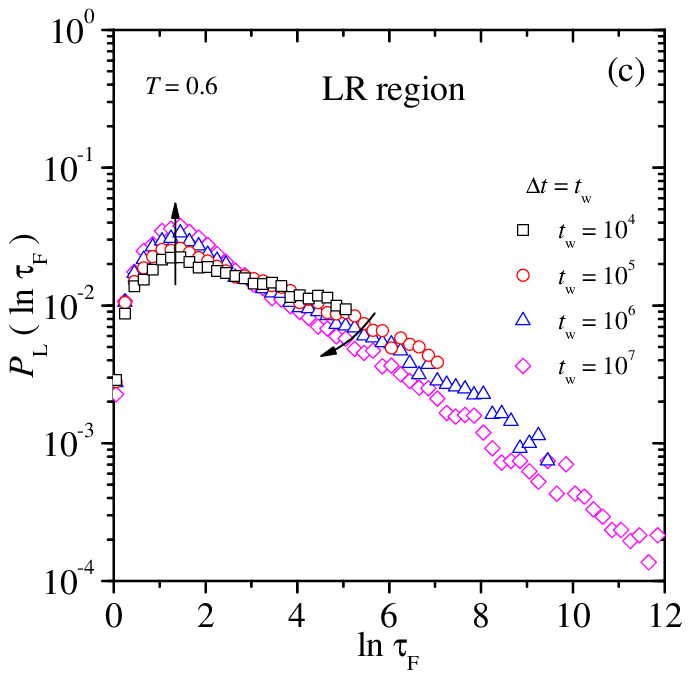}
\caption{\label{fig9} The mean flipping time distributions
for $T=0.6$ and different waiting times, keeping $\Delta t=t_{\mathrm{w}}$. 
Panels show (a) $P$, (b) $P_{\mathrm{H}}$ and (c) $P_{\mathrm{L}}$.  
The arrows emphasize how such distributions evolve with increasing $t_{\mathrm{w}}$.}
\end{center}
\end{figure}

Next, we analyze what happens at constant temperature, $T=0.6$, with increasing $t_{\mathrm{w}}$.  
For simplicity we only show data for $\Delta t=t_{\mathrm{w}}$. 
As expected, the overall behavior is qualitatively rather similar 
to that found in the 3D EAB model (see Fig.~4 in Ref.~\cite{Roma2010a}).  
The full mean flipping time distribution, Fig.~\ref{fig9} (a), 
approximately preserves its shape as the waiting time increases,
although inside and outside the backbone the dynamics is very different.
Figure~\ref{fig9} (b) shows that $P_{\mathrm{H}}$ is as wide as $P$ but evolves slowly in time, 
particularly decreasing for small values of $\tau_{\mathrm{F}}$ (see arrow in this figure).          
In other words, eventually some fast spins tend to slow down its flipping rate.   
On the contrary, Fig.~\ref{fig9} (c) shows that the mean flipping time distribution 
for the LR region evolves in the opposite direction, i. e., with increasing $t_{\mathrm{w}}$
a fraction of the slow spins increases its activity.   
In Ref.~\cite{Roma2010a}, this singular behavior was detected 
in the 3D EAB model and, by comparing with that observed in 
the 3D ferromagnetic Ising model, was interpreted as an indirect evidence 
of the existence of a domain-growth process taking place inside the backbone.   

\section{Discussion and conclusions}

In this work we have analyzed different aspects of the 
nonequilibrium behavior of the 3D EAG spin glass model.  
In particular, we have focused on establishing a link between the dynamical 
and spatial heterogeneities observed on this system.  

For each disorder sample realization, spatial heterogeneities 
are well characterized by the RS, 
a lattice where each bond $J_{ij}$ has been replaced 
by its rigidity $r_{ij}$, a quantity that measures the strengh
of the effective interaction between the spins $\sigma_i$ and $\sigma_j$.  
In the EAB model, the rigidity distribution $F(r)$ is discrete 
and therefore it allows to separate the system in few components,
mainly a backbone and its complement.  
Previous studies suggest that this backbone 
is capable of sustaining a ferromagnetic-like order,
while the rest of the system remain in a paramagnetic phase, 
even below the critical temperature 
\cite{Roma2006,Roma2007a,Roma2007b,Roma2010a,Roma2010b,Rubio2010a,Rubio2010b}.
Remarkably, although the EAG model has a continuous rigidity distribution,
we can determine an adequate 
rigidity threshold, $r_{\mathrm{min}}=1.3$, to separate the
system in a backbone and its complement, 
both having similar topological properties to those found in the bimodal case \cite{Roma2013}.

As considered in Ref.~\cite{Baity-Jesi2014}, a different approach could be to study the 
dynamical heterogeneities by partitioning the rigidity distribution
in many sectors.
Using this idea in the 3D EAB model, which has a discrete rigidity distribution 
with four sectors (corresponding to peaks at $r=0$, $r=4$, $r=8$, and $r=12$), 
no significant differences in the
physical properties between sectors with $r \ge 4$ are observed.
By performing a series of preliminary simulations, 
we have concluded that this is also true for the EAG model,
where we only found qualitative differences when we separate the system into two components.

In Sec.~\ref{NumRes}, we first studied the correlation and the integrated response functions.
While both quantities show a similar trend for the full and the HR region,
for the LR region the curves evolve (with increasing waiting time)
in opposite directions.  In particular, the FDT plots for the latter sector, 
seems to reinforce the idea that the complement of backbone is paramagnetic 
at any nonzero finite temperature. 
If, as proposed in Ref.~\cite{Roma2007b} for the 3D EAB model, 
the backbone is capable of sustaining a ferromagnetic-like order,
then we should observe that the FDT plots tend to a plateau for increasing $t_{\mathrm{w}}$. 
Since the EAG model has a continuous spectrum of energy levels,
implying that the rigidity distribution $F(r)$ is also continuous [see Fig.~\ref{fig1}(b)],
this behavior is probably impossible to observe in such a system
within the longer time scales and the larger lattice sizes which can be simulated.    
 
Finally, we analyzed the mean flipping time distribution.
For the EAG model $P(\ln \tau_{\mathrm{F}})$ is very broad but, 
unlike the EAB model, does not have a two-peaked structure.
Nevertheless, the contributions of the HR and the LR regions
resembles the behavior found in the bimodal system.
Namely, $P_{\mathrm{H}}$ has a single peak that depends on 
temperature as $\ln \tau_{\mathrm{F,max}} \sim 3.8/T$ 
($\sim 4/T$ for the backbone of the EAB model), 
in accordance with an activation process with a 
characteristic mean energy barrier of $\varepsilon \sim 3.8$.
On other hand, $P_{\mathrm{L}}$ also has a single peak 
that depend on $1/T$ but with a smaller slope, $\varepsilon \sim 0.7$.  
Although this effective energy barrier is nonzero
(for the EAB model, the complement of the backbone does not depends on $T$),
notice that it is an order of magnitude smaller than 
the one corresponding to the HR region. 
  
The comparison between the nonequilibrium dynamics and 
spatio-temporal heterogeneities in the EAB and EAG models 
has to be considered with some caution. This comparison, 
within the backbone picture, is based on the assumption that the bond rigidity $r_{ij}$
is a measure of the true interaction strength (effective interaction)
between the pair of spins at sites $i$ and $j$. 
Although both are disordered models, the EAB has a discrete rigidity distribution $F(r)$ 
with only two significant peaks at $r=0$ (the LR region) 
and $r=4$ (this is approximately the HR region; 
in fact, the backbone is formed by bonds with $r \ge 4$ but 
the peaks with $r>4$ are not relevant, see Fig.~\ref{fig1}(a)) \cite{Roma2013}.
As a consequence, this system shows a marked separation between 
paramagnetic-like and ferromagnetic-like phases,
a phenomenon that is most dramatically revealed in the
two-peaked structure of the mean flipping time distribution \cite{Ricci2000,Roma2006,Roma2010a}.
Instead, for the EAG model, the time scale heterogeneous 
behavior is revealed by a very broad rigidity distribution 
$F(r)$ (notice that the $\ln \tau_F$ scale in Fig.~\ref{fig7} 
corresponds to a spread of flipping times within five orders of magnitud) 
and then this phase separation 
is more subtle and difficult to observe.

In conclusion, we find that the nonequilibrium dynamics of the 3D EAG and EAB models,
displays several similitudes when both systems are separated into
their main components, the backbone and its complement. 
For the Gaussian case and below the critical temperature, 
this latter region shows evident signs of a paramagnetic-like behavior.
Instead, within the backbone our data, as for the EAB model, points to the existence of a very slow domain-growth process, most probably with a ferromagnetic-like character.

\section*{Author contribution statement}

All authors designed the study, analyzed the results, wrote and revised the manuscript. F. R. performed the numerical simulations.
\\

\noindent F. Rom\'a acknowledges financial support from CONICET (Argentina) under project No. 
PIP 112-201301-00049-CO, FONCyT (Argentina) under project PICT-2013-0214, 
and Universidad Nacional de San Luis (Argentina) under project PROICO P-31216. 
S. Bustingorry acknowledges partial support by CONICET (Argentina) under Project No.
PIP 112-201201-00250-CO.

\section*{Appendix: Dynamic behavior for different values of $r_{\mathrm{min}}$}
 
\begin{figure}[t!]
\begin{center}
\includegraphics[width=7.3cm,clip=true]{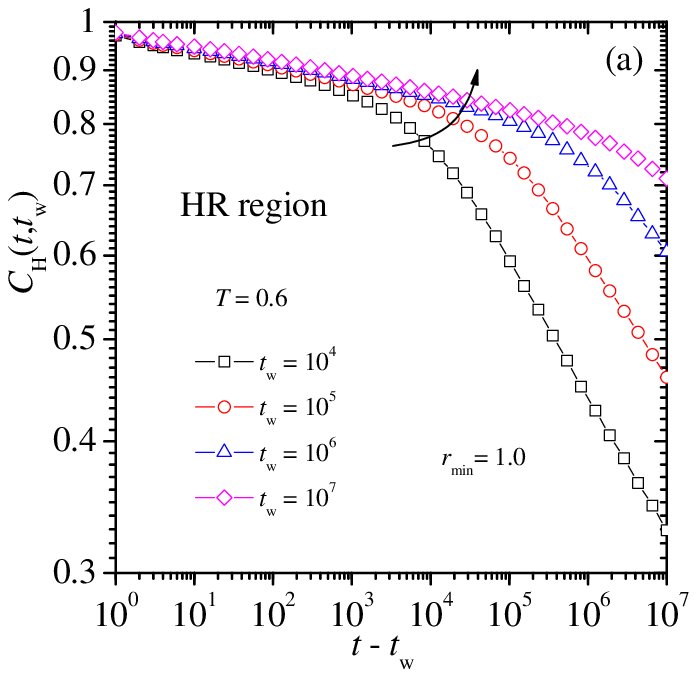}
\includegraphics[width=7.3cm,clip=true]{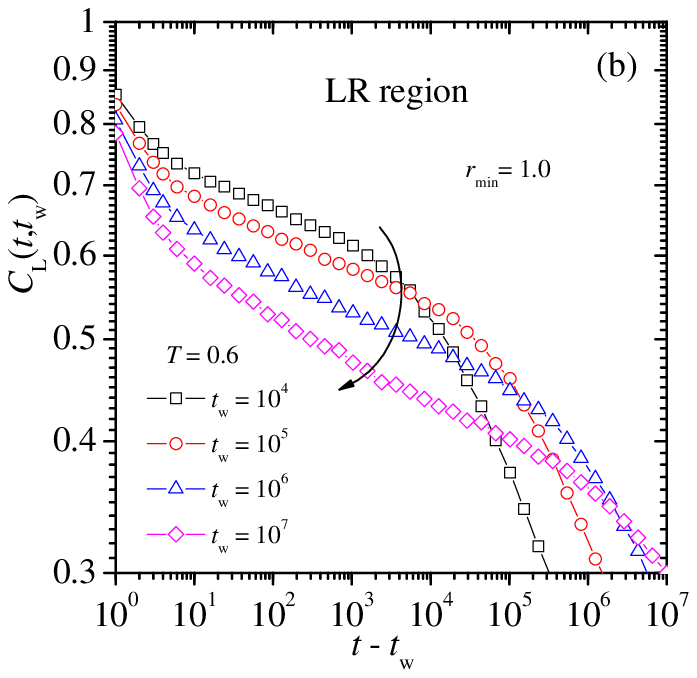}
\caption{\label{figA1} The correlation functions
(a) $C_{\mathrm{H}}$ and (b) $C_{\mathrm{L}}$ at $T=0.6$ and 
for different waiting times as indicated.  
The HR and LR regions have been delimited choosing $r_{\mathrm{min}}=1.0$.  
The arrows emphasize how such functions evolve with increasing $t_{\mathrm{w}}$.}
\end{center}
\end{figure}

For the 3D EAG model, we have selected a rigidity threshold value of $r_{\mathrm{min}}=1.3$ that
separates the lattice into HR and LR regions.
When comparing the number of spins within each region, they look somewhat different:  
the average fraction of spins within the HR and LR regions are $p \approx 0.91$ and $1-p \approx 0.09$, respectively,
suggesting that the backbone is much larger than its complement.
However, note that the backbone is defined in terms of rigidity, which is a bond's property.
For the 3D EAG model, in average, the fraction of bonds within the HR and LR regions are
$h \approx 0.64$ and $1-h \approx 0.36$, respectively, and in terms of bond's fraction, 
the LR region comprises half the size of the HR region.
In Ref.~\cite{Roma2013} it was shown that both regions percolate over the system, 
and thus they are composed of at least
one cluster whose size is of the order of $N$.

\begin{table}[b!]
\caption{Approximate values of the bonds fraction $h$ and the spins fraction $p$ for different values of $r_{\mathrm{min}}$.}
\label{tab:1} 
\begin{center}     
\begin{tabular}{lll}
\hline\noalign{\smallskip}
$r_{\mathrm{min}}$    &   $h$  &   $p$     \\
\noalign{\smallskip}\hline\noalign{\smallskip}
$0.1$                 & $0.97$ &   $1.00$  \\
$0.5$                 & $0.86$ &   $0.98$  \\
$1.0$                 & $0.72$ &   $0.94$  \\
$1.3$                 & $0.64$ &   $0.91$  \\
$2.0$                 & $0.48$ &   $0.81$  \\
$3.0$                 & $0.29$ &   $0.62$  \\
$4.0$                 & $0.16$ &   $0.41$  \\
\noalign{\smallskip}\hline
\end{tabular}
\end{center}     
\end{table}

\begin{figure}[t!]
\begin{center}
\includegraphics[width=7.3cm,clip=true]{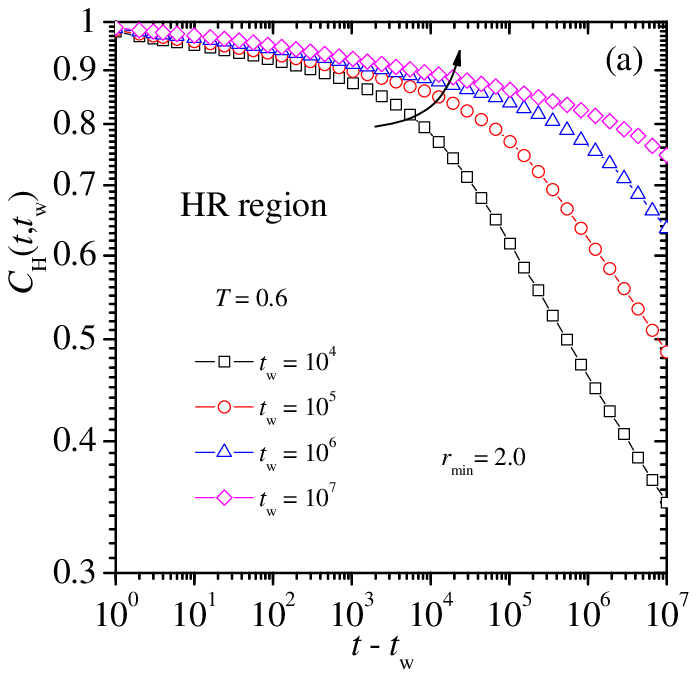}
\includegraphics[width=7.3cm,clip=true]{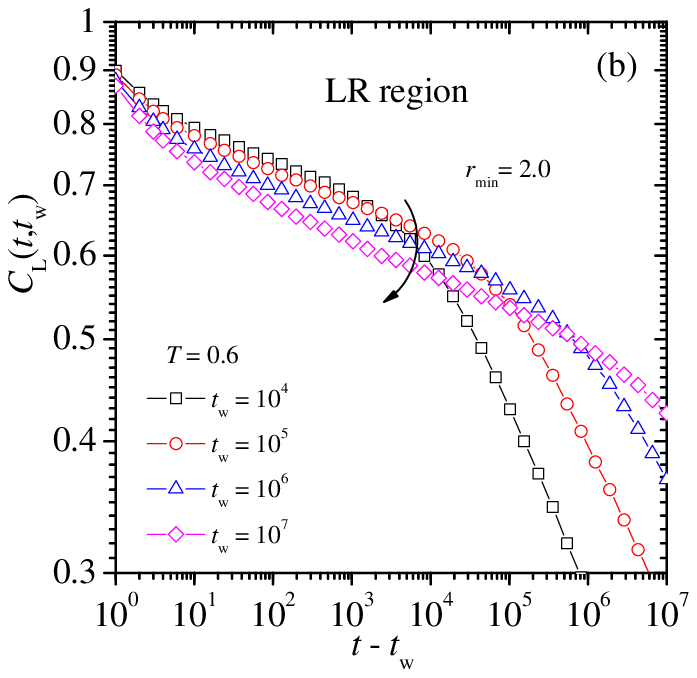}
\caption{\label{figA2} The same as for Figs.~\ref{figA1} (a) and (b) 
but now for $r_{\mathrm{min}}=2.0$.}
\end{center}
\end{figure}

When varying the value of $r_{\mathrm{min}}$ the fractions $h$ and $p$ change, as shown in Table~\ref{tab:1}.
However, the qualitative behavior of the spins belonging to the HR and LR regions do not change when $r_{\mathrm{min}}$
is in the range $0.6 < r_{\mathrm{min}} < 2.0$. 
As an example, we show in Figs.~\ref{figA1} (a) and (b) that for $r_{\mathrm{min}}=1.0$,
the correlation functions $C_{\mathrm{H}}$ and $C_{\mathrm{L}}$
do not suffer significant changes [compare these figures with Figs.~\ref{fig3} (b) and (c)].
Also, we observe a similar trend if we take $r_{\mathrm{min}}=2.0$, Figs.~\ref{figA2} (a) and (b).     
For values of $r_{\mathrm{min}} > 2$, both $C_{\mathrm{H}}$ and $C_{\mathrm{L}}$ 
behave in the same way that the full correlation functions $C$ 
(for simplicity these curves are not shown here).
On the other hand, for $r_{\mathrm{min}} < 0.6$ (below the range $[0.6-2.0]$),
we can still observe curves like those shown in Figs.~\ref{figA1} and \ref{figA2} 
but in this case, where the size of the LR region is very small (see Table~\ref{tab:1}), 
the behavior of $C_{\mathrm{L}}$ corresponds to a few 
quasi-free spins (spins that can flip with a negligible energy cost)
which represent only a fraction of the true (larger) LR region.   
The remaining observables show a similar behavior
with respect to changes in the value of $r_{\mathrm{min}}$.

%.....................................................................

\end{document}